\documentclass[aps,floats,superscriptaddress,twocolumn,pre]{revtex4}
\usepackage{amssymb,amsmath}
\usepackage{amsmath,amssymb}
\usepackage{graphicx}
\usepackage{psfrag,textcomp}

\def\beq{\begin{equation}}
\def\eeq{\end{equation}}
\def\bea{\begin{eqnarray}}
\def\eea{\end{eqnarray}}
\begin{document}
\title{Phase coexistences and particle non-conservation in a closed
asymmetric exclusion process with inhomogeneities}
\author{Tirthankar Banerjee}\email{tirthankar.banerjee@saha.ac.in}
\affiliation{Condensed Matter Physics Division, Saha Institute of
Nuclear Physics, Calcutta 700064, India}
\author{Anjan Kumar Chandra}\email{anjan.chandra@saha.ac.in}
\affiliation{Condensed Matter Physics Division, Saha Institute of
Nuclear Physics, Calcutta 700064, India}
\affiliation{Department of Physics, Malda College, Malda, India}
\author{Abhik Basu}\email{abhik.basu@saha.ac.in}
\affiliation{Condensed Matter Physics Division, Saha Institute of
Nuclear Physics, Calcutta 700064, India}
\date{\today}

\begin{abstract}
We construct a one-dimensional totally asymmetric simple exclusion process (TASEP) on a ring with two segments having unequal hopping rates,
coupled to particle non-conserving Langmuir kinetics (LK) characterized by equal attachment and detachment rates. In the steady state, in the limit of competing LK and TASEP, the model is always found in
states of phase coexistence. We uncover a nonequilibrium phase transition  between a
  three-phase and a  two-phase coexistence in the faster segment, controlled by the
 underlying inhomogeneity configurations and LK. The model is always found to be half-filled on average in the steady state, regardless of the hopping rates and the attachment/detachment rate.

\end{abstract}

\maketitle

\section{Introduction}
Totally asymmetric simple exclusion process (TASEP) and its variants with open boundaries in one dimension (1D) serve as simple models of restricted 1D transport. These 1D transports are observed in a variety of situations, e.g., motion in nuclear pore complex of
cells~\cite{nuclearpore}, motion of molecular motors along microtubules~\cite{molmot}, fluid flow in artificial
crystalline zeolites~\cite{zeo} and protein synthesis by messenger
RNA (mRNA) ribosome complex in cells~\cite{albertbook}; see, Refs.~\cite{review} for basic reviews
on asymmetric exclusion processes.
The coupled dynamics of TASEP and random attachment-detachment in the form of Langmuir kinetics (LK) displays a rich behavior
including coexistence of low and high density regions of particles and a boundary condition independent phase,
in the limit when LK competes with TASEP~\cite{Frey-LK}. Open TASEPs with defects, both point and extended,
have been studied; see, e.g. Refs.~\cite{tasep-def} which investigated the effects of the defects on the steady state densities and currents.
In addition, open TASEP with a single point defect along with LK has been considered in Ref.~\cite{erwin2},
which finds a variety of phases and phase coexistences as a result of the competition between the defect and LK.

In recent studies involving asymmetric exclusion processes on  closed inhomogeneous rings, the total particle number is held fixed by the dynamics, as expected in
exclusion processes; see, e.g., Refs.~\cite{Mustansir1,niladri,tirtha1}. Nonconserving LK is expected to modify the steady
state densities of pure TASEP on a closed inhomogeneous ring. TASEP on a
perfectly homogeneous ring yields uniform steady state densities, due to the
translational invariance of such a system. Evidently, introduction of the
particle nonconserving LK should still yield uniform steady state
densities, again due to the translational invariance of the system, although the actual
value of the uniform steady state density should now depend upon LK.
Nonuniform or inhomogeneous steady states are expected only with explicit breakdown of the translation invariance, e.g., by means of quench disorder
in the hopping rates at different sites. Studies on this model should be useful  in various contexts, ranging from vehicular/pedestrial traffic to
ribosome translocations along mRNA, apart from theoretical interests. For example, consider pedestrian or vehicular movement along a circular track with
bottlenecks/constrictions,
where overtaking is prohibited and pedestrians or vehicles can either leave and join the circular
track (say, through side roads) randomly~\cite{review}, or, for instance, consider the motion of ribosomes along closed mRNA loops with defects
where the ribosomes can attach/detach to the mRNA loop stochastically~\cite{mrna}.

 In this article, we introduce a disordered TASEP on a ring with LK,
 where the disorder is in the form of piecewise discontinuous hopping rates across the two segments of the ring.  The unidirectional hopping of the particles across a slow segment yields reduced particle current. Evidently, this breaks the
translational invariance. Hence, inhomogeneous steady state densities cannot be
ruled out. In addition, we allow random attachment-detachment of the particles or LK at every site of the ring.  Thus, the interplay of the quenched disorder in the hopping rate with the consequent
absence of translation invariance and LK should determine the steady states of the model. For simplicity we assume equal rates for attachment and detachments. Our model is well-suited to analyze a key question of significance, {\em viz}, whether the steady state density profiles and the average particle numbers in the steady states can be controlled by the disorder and (or) the LK.
Recent studies of nonequilibrium steady states in TASEP on a ring with quenched disordered
hopping rates without any LK revealed macroscopically inhomogeneous steady state densities in
the form of a localized domain wall (LDW) for moderate average particle densities in the system;
see, e.g., Refs.~\cite{Mustansir1,lebo}.
Our work provides insight about how the steady states
in the models in Refs.~\cite{Mustansir1,lebo} are affected by particle nonconservation and allows
us to study competition between bulk LK and asymmetric exclusion processes in ring geometry. To our
knowledge, this has not been studied before. We generically find (i) nonuniform steady states and phase coexistences, (ii) phase transition between   different states of phase coexistences and (iii) the system is always half-filled in the steady state for the whole relevant parameter range, regardless of the detailed nature of the underlying steady state density profiles. The rest of the article is organized as follows. In Sec.~\ref{one}, we construct our model.
Then we calculate the steady state density profiles for an extended defect in Sec.~\ref{1a} and for a point
defect in Sec.~\ref{model2}. In Sec.~\ref{comparison}, we compare the results for
extended and point defects. Next, in Sec.~\ref{smallomega}, we discuss why the average density shows a fixed value for any choice of
the phase parameters. Finally, in Sec.~\ref{conclu} we summarize and conclude.

\section{The Model}\label{one}
\begin{figure}[htb]
\includegraphics[height=6.5cm,width=6.0cm]{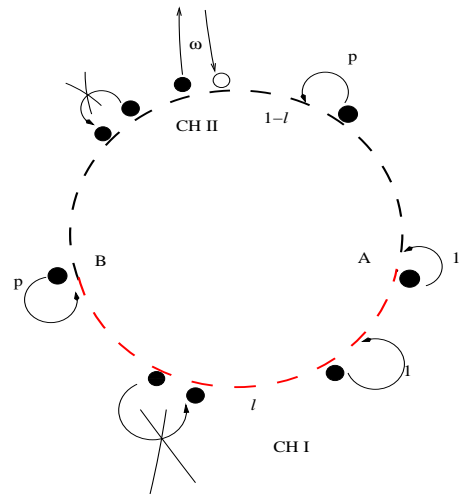}
\caption{(Color online) Particles hop with rates $1$ and $p<1$ in segments CHI (red)
and CHII, respectively. $\omega=\Omega/N$ (see text) denotes both the evaporation and
condensation rates; A and B are the junctions where the two segments meet.} \label{model}
\end{figure}

We consider an exclusion process on a closed 1D inhomogeneous ring
with $N$ sites, together with nonconserving LK. The quenched
inhomogeneity is introduced via space-dependent hopping rates. The parts with lower
hopping rates can be viewed as defects in the system.   Specifically, our model consists of
 two segments of generally unequal number of sites. We call the parts
as channel I (CHI) with $N_1$ sites (sites $i=1,2,...,M$) and unit hopping rate, and channel II (CHII) with $N_2$ sites (sites $i=M+1,M+2,...,N$) and hopping rate $p<1$, where $M<N$
(see Fig.~\ref{model}). We consider both the cases of an extended and a point defect separately.
The size of an extended defect scales with the system size, such that even in the thermodynamic limit,
it covers a finite fraction of the ring (i.e., finite $N_2/N$ in the limit $N\rightarrow \infty$).
In contrast, the size of a point defect does not scale with the system size, and hence,
the size of a point defect relative to the system size vanishes
in the thermodynamic limit (i.e., $N_2/N\rightarrow 0$ for $N\rightarrow\infty$). Now consider an asymmetric exclusion process on the ring: A
particle can hop in the anticlockwise direction to its neighboring site iff the site is empty. No two
particles can occupy the same site and neither can a particle move
backward even if there is a vacancy.
 Between CHI and CHII, particle
exchanges are defined at the junctions A and B only (see Fig.~\ref{model}), where dynamic rules
are defined by the originating site. In addition, the system executes LK, e.g., at any site, a particle can either
attach to a vacant site  from the surroundings or leave an occupied
site at a rate $\omega$.

%

 With $n_{Ii}$ and $n_{IIi}$ as
the steady state number densities at $i$-th sites of CHI and CHII,
respectively and $N_{tot}$  the total number of particles
in the system ($N_1=lN, N_2=(1-l)N$, $l$ refers to the fraction of sites
having hopping rate unity),
\begin{equation}
 [n_{I}l+n_{II}(1-l)]N=N_{tot} ,
\end{equation}
where, $n_I=\sum_in_{Ii}$, $n_{II}=\sum_i n_{IIi}$.
Now define the mean number density for the total system, $n$ as
\begin{equation}
n=\frac{N_{tot}}{N}={n_{I}l+n_{II}}(1-l). \label{ratio1}
\end{equation}
Due to  LK, $n$ is not a conserved quantity and cannot be used to characterize the steady states in the model,
unlike~\cite{Mustansir1}. Rather,
$l$, $p$ and $\omega$  parametrize the steady density profiles.
In order to ensure that the total flux of the particles due to LK is comparable to the particle current due to the hopping dynamics of TASEP (i.e., the total detachment-attachment
events of the particles due to LK
should be comparable to the number of crossings of the
junctions A and B by the particles in a given time interval),
we introduce a scaling for the evaporation/condensation  rates and
define the total rate, $\Omega=\omega N$ and analyze the system for a given
$\Omega\sim O(1)$. This ensures competition between LK and TASEP; see, e.g. Ref.~\cite{Frey-LK}.
Although there is no particle number conservation either in Ref.~\cite{Frey-LK} or here,
it is important to emphasize one important difference between the two that stems
from the fact that our model is closed. As a result, there is no injection or extraction
of particles at designated "entry" or "exit" sites unlike in open TASEP or the model in Ref.~\cite{Frey-LK}, where these rates are the tuning parameters.

\section{Steady state densities}
We perform  Mean field (MF) analysis of our model, supplemented by  its extensive Monte Carlo Simulations (MCS).
\subsection{MF analysis and MCS results for an extended defect}\label{1a}

Before we discuss the details of the MF analysis of our model, it is
useful to recall the results from the model in Ref.~\cite{Frey-LK} where
the steady state densities of a TASEP with open boundaries together
with LK having equal attachment and detachment rates, $\Omega$ are
investigated. Depending upon the entry ($\alpha$) and exit ($\beta$)
rates and $\Omega$, the steady state densities can be low or high
densities, or phase coexistences involving three or two phases. By
varying the above control parameters, transitions between the
different steady states are observed. Motivated by these results
and considering the junctions between the two segments as the
effective entry and exit points of the segments, it is reasonable to
expect similar behavior including phase coexistences and transitions
between them in our model. Our detailed analysis as given below
partly validate these expectations; we show that our model displays phase 
coexistences, but has no analogs of the low and high density phases. We find 
that even in the case of a point
defect where there is effectively only one junction, these remain
true.

Our MF analysis is based on treating the model as a combination of two TASEPs - CHI and CHII,  joined at the junctions A and B, respectively. Thus, junctions B and A are {\em effective}
entry and exit ends of CHI. This consideration allows us to analyze the
phases of the system in terms of the known phases of the open boundary LK-TASEP~\cite{Frey-LK}.  For
convenience, we label the sites by a continuous variable $x$ in the
thermodynamic limit, defined by $x=i/N$, $0<x<1$. In terms of the rescaled coordinate $x$, the lengths of
CHI and CHII are $l$ and $1-l$, respectively. For an extended defect here, $l<1$. Without the LK dynamics,
the steady state densities of a TASEP on an inhomogeneous ring may be
obtained by means of the conservation of the total particle number and the
particle current in the system~\cite{Mustansir1,niladri,tirtha1}. In contrast,
it is important to note that in the present model, due
to the nonconserving LK dynamics, the particle current is conserved only {\em locally},
since the probability of attachment or
detachment at a particular site vanishes as $1/N$~\cite{Frey-LK}.
Similar to Ref.~\cite{Frey-LK},  the steady state densities, $n_I(x),\,n_{II}(x)$ follow


\begin{eqnarray}
 (2n_I - 1)\left(\frac{dn_I}{dx}- \Omega\right)&=&0,\label{eqbasic1}\\
 (2n_{II}-1)\left(\frac{dn_{II}}{dx}-\Omega\right)&=&0.\label{eqbasic}
\end{eqnarray}
These yield $n_I,\,n_{II}=1/2$ and $n_I,\,n_{II}=\Omega
x+C_I/C_{II}$, where $C_I,C_{II}$
are the integration constants.

Apply now the current conservation locally at A and B.
Ignoring possible boundary layers,  this
yields
\begin{equation}
 n_I(1-n_I)=pn_{II}(1-n_{II}) \label{currcon}
\end{equation}
separately, very close to $x=0$ and $x=l$. Since $p<1$ necessarily, Eq.~(\ref{currcon}) yields $n_I\neq 1/2$. Thus,
either $n_I>n_{II}$ or $n_I <n_{II}$ very close to the junctions A and B. It is
known that with $\Omega=0$, $n_{II}=1/2$ is a solution for moderate $n$, with $n_I(x)$ being in the form of a localized domain wall (LDW)~\cite{Mustansir1, tirtha1}.
Finite $\Omega$ is expected to modify these solutions. Nonetheless, since
$n_{II}=1/2$ is a solution of the steady state equation (\ref{eqbasic}),  $n_{II}=1/2$ remains a valid steady
state solution for non-zero $\Omega$.  Whether or not there are
other solutions for $n_{II}$ is discussed later.  With $n_{II}=1/2$, we obtain $n_I(x)$ at $x=0,l$ by  (\ref{currcon}).

Application of (\ref{currcon}) yields
\begin{equation}
n_I=\frac{1 \pm \sqrt{1-p}}{2},\label{n1val}
\end{equation}
at $x=0,l$, which serve as boundary conditions on $n_I(x)$. It is useful to compare CHI with an open TASEP.
We identify effective entry ($\alpha_e$) and exit ($\beta_e$) rates: $\alpha_e = \beta_e = {(1 - \sqrt{1-p})}/{2}\leq 1/2$.
Considering
the fact that the hopping rate $p$ of CHII is less than that in CHI (unit
value), on
physical grounds,
we expect particles to accumulate behind junction A in CHI only. In other
words, we expect $n_I(x=l)\geq n_I(x=0)$. These considerations allow us to
set the boundary conditions $n_I(x=0)=[1-\sqrt{1-p}]/2$ and
$n_I(x=l)=[1+\sqrt{1-p}]/2$.
 Hence from Eq.(\ref{eqbasic1}), we arrive at the three following
solutions for $n_I(x)$, namely
\begin{equation}
 n_{I\alpha}(x)=\Omega x + \frac{1-\sqrt{1-p}}{2}
\end{equation}
\begin{equation}
 n_{I\beta}(x)=\Omega(x-l) + \frac{1+\sqrt{1-p}}{2}
\end{equation}
and
\begin{equation}
 n_{Ib}=\frac{1}{2}
\end{equation}
{ where $n_{I\alpha}(x)$ and $n_{I\beta}(x)$ are the linear density profiles satisfying the boundary conditions
at the entrance (B) and exit (A) ends of CHI and $n_{Ib}$ represents the MC region.
Notice that  the solution $n_{Ib}=1/2$ cannot be
extended to the junctions A and B, else (\ref{currcon}) will be violated.

Given the physical expectation that  $n_I(x)$ should not decrease with $x$,
we identify two values of $x$, viz., $x_\alpha$
and $x_\beta$ where the linear solutions meet with the third solution.
Depending on these values of $x_\alpha$ and $x_\beta$, we will see
that the system is found in various phases which are parametrized by
$p$ and $\Omega$. Using $n_{I\alpha}(x_\alpha)=n_{I\beta}(x_\beta)=
\frac{1}{2}$, we get $x_\alpha=\frac{\sqrt{1-p}}{2\Omega}$ and $x_\beta
=l-\frac{\sqrt{1-p}}{2\Omega}=l-x_\alpha$.
Thus we find
\begin{equation}\label{main eq}
 x_\beta-x_\alpha=l-\frac{\sqrt{1-p}}{\Omega}.
\end{equation}
Hence there are three distinct possibilities, namely
$x_\alpha=x_\beta$, $x_\alpha < x_\beta$, $x_\alpha > x_\beta$, depending on
which the system will be found in different phases. We now analyze each
of the cases in detail.
\par
$(1)$ Consider $x_\alpha < x_\beta$. Here we observe a three-phase coexistence. Near $x=0$, we
see a low density (LD) phase  having density, $n_{Ix}<1/2$, rising with a
positive slope upto $x=x_\alpha$. For $x_\alpha<x<x_\beta$, there is a maximal
current (MC) phase with $n_{Ix}=1/2$ and the current, $J_{Ix}=1/4$ and while
$x_\beta<x<l$, we see a high density (HD) phase with $n_{Ix}>1/2$. This is
accompanied by an MC phase in CHII. Representative plot of comparisons of
MFT and MC results are shown in Fig.~\ref{3_phase}.
\par
$(2)$ When $x_\alpha=x_\beta$, the maximal current region separating the two linear
solutions vanishes and the density profile becomes an inclined straight
line, matching continuously the densities of the LD and HD phases, see Fig.~\ref{straight_line}.
\par
$(3)$ As $x_\alpha > x_\beta$, we can no more find the MC region and instead find
a density discontinuity. Solutions from the left and right meet at a point $x_w$
in the bulk of CHI in the form of a localized domain wall (LDW), where the left and right currents, i.e., $J_\alpha (x_w)$ and
$J_\beta (x_w)$ are equal. We can arrive at an expression for $x_w$ using the
the local current conservations. Here, $J_\alpha(x_w)=n_{I\alpha}(x_w)\left(1-n_{I\alpha}(x_w)\right)$.
Similarly, $J_\beta(x_w)=n_{I\beta}(x_w)\left(1-n_{I\beta}(x_w)\right)$. The equality of
$J_\alpha(x_w)$ and $J_\beta(x_w)$ gives us the condition
\begin{equation}\label{dw eq}
 n_{I\alpha}(x_w)+n_{I\beta}(x_w)=1,
\end{equation}
since $n_{I\alpha}(x_w) \neq n_{I\beta}(x_w)$. Using Eq.(\ref{dw eq}), $x_w= l/2$.
Thus, the LDW is always at the midpoint of CHI, unlike in Refs.~\cite{niladri,tirtha1}.
The fact that $x_w=l/2$ may be understood from the symmetrical structures of $n_{I\alpha}(x)$
and $n_{I\beta}(x)$. Notice that $1/2-n_{I\alpha}(x=0)=n_{I\beta}(x=l)-1/2$. Since, the upward
slope of $n_{I\alpha}(x)$ is same as the downward slope of $n_{I\beta}(x)$, which is $\Omega$,
equality of the currents $J_\alpha (x_w)=J_\beta (x_w)$ ensures that $x_w=l/2$. Thus, $x_w$ is
{\em independent of} the values of $\Omega$ and $p$. This is to be contrasted with Ref.~\cite{Frey-LK}
where $\Omega$ is known to affect the location of the LDW there. The height
of the LDW is
\begin{equation}
\Delta h_e=n_{I\beta}(x_w)-n_{I\alpha}(x_w)=\sqrt{1-p}-\Omega l,\label{htextnd}
\end{equation}
which depends upon $l$, $p$ and
$\Omega$. In our MCS studies we have generally chosen $l=1/2$ without any loss of generality.
Since $l=1/2$, the domain wall in CHI will always be located at $x_w=1/4$, irrespective
of the values of $p$ and $\Omega$.
See Fig.~\ref{dw_profile} for a representative plot. The overall average density $n$ may be found from (neglecting the boundary layers)
\begin{equation}
 n=\int_0^l n_I (x) dx + \int_l^1 n_{II}(x) dx.
 \end{equation}
Substituting the MF forms of $n_I(x)$ and $n_{II}(x)$, it is clear that $n=1/2$, regardless of whether CHI is in its
two- or three-phase coexistence state. Our MCS results agree with the MF results to a good extent.

\begin{figure}[htb]
\includegraphics[height=6.5cm,width=8.3cm]{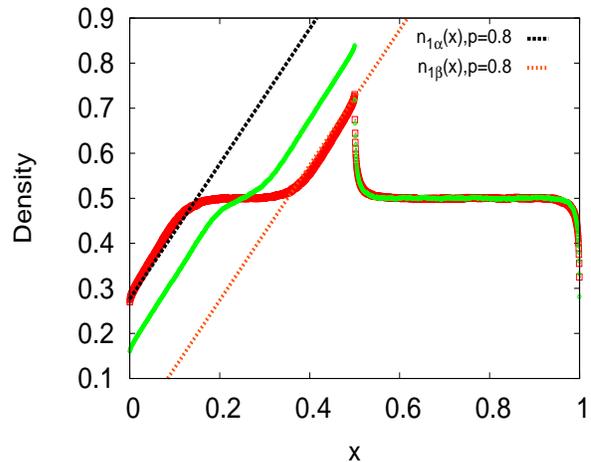}
\caption{(Color online) Steady state density profiles showing three-phase coexistence in CHI for $l=0.5, N=2000$.
The left and right solutions meet at $n(x)=1/2$ in the bulk. Square points represent MCS
results for $p=0.8,\Omega=1.5$; while the deep and shaded lines show the corresponding
MFT equations for $n_{1\alpha}$ and $n_{1\beta}$. The circular points represent MCS studies for
$p=0.55, \Omega=1.5$. Clearly, the length of the MC region varies depending on the parameters, $p$
and $\Omega$.} \label{3_phase}
\end{figure}

\begin{figure}[htb]
\includegraphics[width=9.0cm]{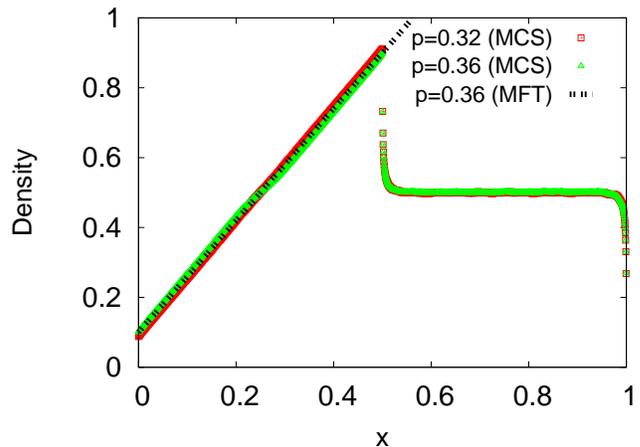}
\caption{(Color online) Straight line density profile for the
extended defect ($x_\alpha=x_\beta,l=0.5, N=2000$) in CHI. The dotted black line represents
the MFT denisty profile equation for $\Omega=1.6,p=0.36$ and the triangular points
shows the corresponding MCS result. The square points correspond to the
MCS density profile for $\Omega=1.6,p=0.32$.} \label{straight_line}
\end{figure}

\begin{figure}[htb]
\includegraphics[width=9.0cm]{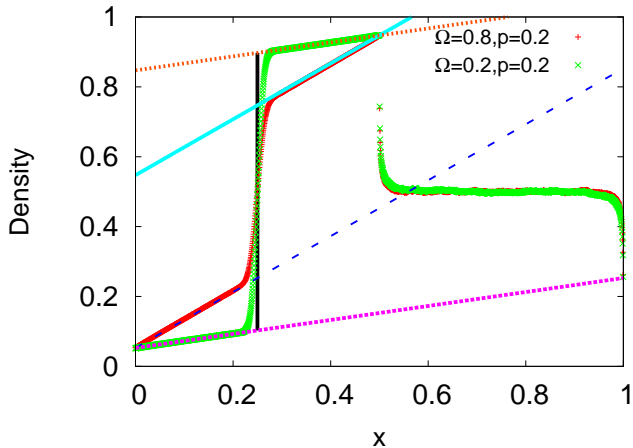}
\caption{(Color online) Formation of an LDW in CHI for $x_\alpha > x_\beta$ with
 $l=0.5, N=2000$. Two different data sets have been plotted for $p$ and $\Omega$.
 Points represent MCS results. The lines represent the following (all MFT equations):
 Broken rectangles($n_{1\beta},\Omega=0.2,p=0.2$), Broken square($n_{1\alpha}, \Omega=0.2,p=0.2$),
 Solid($n_{1\beta},\Omega=0.8,p=0.2$), Gapped ($n_{1\alpha},\Omega=0.8,p=0.2$). As expected,
 the location of the domain wall remains fixed (continuous line at $x=0.25$ represents the MFT DW).} \label{dw_profile}
\end{figure}

We now discuss why the system cannot be found in any other combination of the phases of an open TASEP. First of all,
there is no possibility of only an MC phase in CHI. This follows from the fact that if $n_I=1/2$,
Eq.~\ref{currcon} would be violated at the boundaries.
In order to have an LDW  in CHII, particles should pile up behind CHI which
is physically unexpected since CHI has a higher hopping rate.
Further, we argue that with an MC phase in CHII, CHI cannot be found in a pure LD phase. For CHI to be in such
a phase, $x_\alpha=1/2$. But it is also necessary that $x_\beta$ has to be greater than
$x_\alpha$, otherwise CHI will show an LDW. But the maximum possible
value for $x_\beta$ is $1/2$. Now, $x_\alpha=x_\beta=1/2$ implies $p=1$, for which the system becomes homogeneous. Since we necessarily have $p<1$ in our model, a pure LD phase for CHI with an MC phase in CHII is
ruled out. Due to the particle-hole symmetry, we rule out a pure HD phase for CHI with
CHII  in its MC phase. Lastly, both CHI and CHII cannot be in their LD phases. This may be understood as follows. The general solutions for Eqs.~(\ref{eqbasic1}) and (\ref{eqbasic}) are either inclined lines of the form $\Omega x + C_{I,II}$ ($C_{I,II}$ being constants) or flat $(1/2)$.
  For LD phases, $C_{I,II}$ are to be determined by the density values at the entry sides of CHI and CHII, respectively.
Let us assume $n_I(x=0)=n_{Il}$ and $n_{II}(x=l)=n_{IIl}$. With these known values,  $n_I(x=l)$ and $n_{II}(x=1)$ respectively in CHI and CHII
can be determined. Say, these values are $n_{Ir}$ and $n_{IIr}$, for the respective
channels. Clearly, $n_{Ir}>n_{Il}$ and $n_{IIr}>n_{IIl}$.
But $n_{Ir}$ is connected to $n_{IIl}$ at junction A by
\begin{equation} \label{ld1}
 n_{Ir}(1-n_{Ir})= pn_{IIl}(1-n_{IIl})
\end{equation}
and $n_{IIr}$ and $n_{Il}$ are connected at junction B by
\begin{equation} \label{ld2}
 pn_{IIr}(1-n_{IIr})=n_{Il}(1-n_{Il}).
\end{equation}
Clearly, both (\ref{ld1}) and (\ref{ld2}) cannot be satisfied, simultaneously. Hence this
rules out the LD-LD phase for the system. Similar arguments rule out simultaneous HD phases in both channels.

\subsubsection{The phase diagram}\label{1b}
We now discuss the phase diagram spanned in the $\Omega$-$p$ space.
Consider the case when $x_\alpha=x_\beta$. Using Eq.~\ref{main eq},
\begin{equation}
 \sqrt{1-p}=\Omega l,
\end{equation}
which gives the phase boundary :
\begin{equation}\label{phase_boundary}
 p=1-\left[\Omega l \right]^{2}
\end{equation}
Thus when $x_\alpha < x_\beta$, we have $p > 1-\left[\Omega l_1 \right]^{2}$
and the system shows three-phase coexistence of LD ($n_I<1/2$), MC ($n_I=1/2$) and HD ($n_I>1/2$) regions in
CHI. For
$p < 1-\left[\Omega l \right]^{2}$, we get an LDW in CHI.
For both three-phase and two-phase coexistences in CHI, CHII will be in its MC phase.
The phase diagram is shown in Fig.\ref{phase_diagram}.


\begin{figure}[htb]
\includegraphics[height=6.5cm]{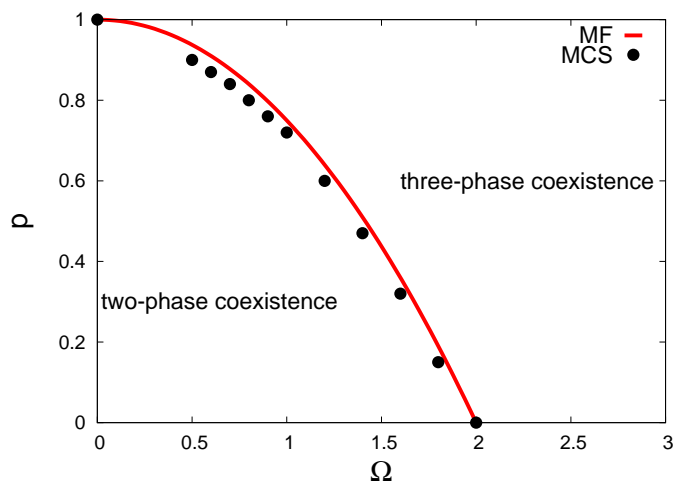}
\caption{(Color online) The phase diagram for $l=0.5$. The curve
$p=1-\frac{\Omega^2}{4}$ separates regions of two-phase
and three-phase coexistences.} \label{phase_diagram}
\end{figure}


The width $W$ of the MC region (numerically equal
to $x_\beta-x_\alpha>0$) in the three-phase coexistence, can be identified as the order parameter
for the phase transition between a three-phase coexistence and a two-phase coexistence.
When the system is in three-phase coexistence,
$W > 0$, where as it is zero in the two-phase coexistence. For a fixed value of $p$, as $\Omega$ is increased,
the system makes a transition from a two-phase coexistence state to a three-phase
coexistence state following Eq.(\ref{phase_boundary}). Accordingly, $W$ increases from
$0$ to $l$ as $\Omega$ is increased for a fixed $p$. 
At the transition, $W=0$. We now find how $W$ approaches zero as $p\rightarrow p_c$
or $\Omega\rightarrow\Omega_c$ in the three-phase coexistence; $p_c,\Omega_c$ are located at the phase boundary
given by (\ref{phase_boundary}). Writing $p=p_c+\delta p,\Omega=\Omega_c+\delta\Omega$ for small
$\delta p>0,\delta\Omega >0$, we have
 \beq
 W=l-\frac{\sqrt{1-p_c}}{\Omega_c}[1-\delta p/2-\delta\Omega]= \frac{\sqrt{1-p_c}}{\Omega_c}(\delta p/2+\delta\Omega),
 \eeq
 to the linear order in $\delta p,\delta\Omega$. Evidently, $ W$ vanishes smoothly as $\delta p$ and $\delta\Omega$,
 indicating the second order nature of the phase transition. Thus, considering either $p$ or $\Omega$ as the
 control parameter (for fixed $\Omega$ or $p$, respectively) and drawing analogy with equilibrium second order phase transitions~\cite{chaikin} we extract an "effective"
 order parameter exponent of value unity. This is to be contrasted with the MF order parameter exponent of value 1/2 in
 equilibrium critical phenomena. This difference is not surprising, considering that the present model is inherently out of equilibrium.

\subsection{Density profiles for a point defect}\label{model2}

Consider now the extreme limit with $l\rightarrow 1$, i.e., with a point defect.  In this limit,
the system has only one site where the hopping rate, $p$ is less than $1$ while
for all other sites, the hopping rate is unity. Thus the MF analysis above by considering the system to be
a combination of two TASEP channels joined at two ends no longer works because CHII (as defined
for an extended defect) now contains just one site and has a vanishing length relative to the whole system for
$N\rightarrow\infty$.
Instead, the system is  just one TASEP, say CHI, with a density $ n_I(x),0<x<1$ and two of its
ends joined at one site having a hopping rate $p<1$.

\par
Let the defect be present at $x=0$ (which is same as $x=1$). We assume that the particles are hopping
anticlockwise as before.
On physical grounds we expect piling up of particles (if at all) should occur behind the blockage site at $x=0$.
\par
Now assume a macroscopically nonuniform steady state density profile, such that there is a pile up of particles
behind the defect at $x=0$ and hence a jump in the density at $x=0$.
Let $\rho_I$ and $\rho_{II}$ be the  densities just to the left and right of $x=0$, respectively: $ n_I(1-\epsilon)=\rho_I,\, n_I(\epsilon)=\rho_{II},\epsilon\rightarrow 0$
Using current conservation at $x=0$, we write
\begin{equation}
 \rho_I(1-\rho_I)=p\rho_I(1-\rho_{II})=\rho_{II}(1-\rho_{II})
\end{equation}
This yields solutions for $\rho_I$ and $\rho_{II}$,viz.,
$\rho_I=\frac{1}{1+p}$ and $\rho_{II}=\frac{p}{1+p}$.
 Density $ n_I(x)$ satisfies the equation
 \beq
 (2 n_I-1)\left(\frac{d n_I}{dx} -\Omega\right)=0,
 \eeq
 yielding
 solutions
\begin{equation}
 n_I(x)=1/2, \Omega x + C.
\end{equation}
The constant of integration $C$ is to be fixed by using either of the boundary conditions $\rho_I$ and $\rho_{II}$. These
evidently yield two values of $C$, say $C_R$ and $C_L$, respectively,  giving $C_L=\frac{p}{1+p}$
and $C_R=\frac{1}{1+p}-\Omega$. Therefore, the two solutions of $ n_I(x)$ are
\begin{eqnarray} \label{pointmf}
 n_{L}(x)=\Omega x+\frac{p}{p+1}\\
 n_{R}(x)=\Omega (x-1)+\frac{1}{p+1},\nonumber
\end{eqnarray}
along with the uniform solution $n_{b}=1/2$.
Similar to an extended defect, we can compare CHI in case of point defect
with an open TASEP and extract effective entry and exit rates: $\alpha_p = \beta_p = p/(p+1)< 1/2$ where
$\alpha_p$ and $\beta_p$ are entry and exit rates respectively.

Since $ n_{L}(x)$ and $ n_{R}(x)$ depend linearly upon $x$, in general they should meet with the uniform solution,
i.e., $n_b=1/2$ at
two points say, $x_{L}$ and $x_{R}$. The quantitative analysis follows the same logic as above for an extended
defect. Accordingly, the values of $x_L$ and $x_R$
will determine whether the system is in its three-phase coexistence state or a two-phase
coexistence state.
We find
\begin{equation}\label{point_L}
 x_L= {\left(1/2-\frac{p}{p+1}\right)}\frac{1}{\Omega},
\end{equation}
\begin{equation}\label{point_R}
 x_R= {\left(1/2+\Omega-\frac{1}{1+p}\right)}\frac{1}{\Omega}.
\end{equation}
The system will thus be in three-phase coexistence when $x_L<x_R$ and we will have
two inclined lines meeting the third solution in the bulk at $x_L$ and $x_R$,
respectively. The extent of the MC phase (bulk solution) is given by
 \beq
x_{R}-x_L=1-\left(\frac{1-p}{1+p}\right)\Omega.
 \eeq
 Similarly, for $x_L>x_R$, the system will be found in its two-phase
coexistence
state and the two inclined solutions will meet in the bulk in the form of an LDW.
The location of the LDW, $x^p_w$ may be calculated similarly as that for an extended defect, yielding $x^p_w=\frac{1}{2}$. The height of the LDW is  density difference between $n_{IL}(x)$ and $n_{IR}(x)$ at $x=1/2$ which is given by
 \begin{equation}
\Delta h_p = n_{R}(x_p)-n_{L}(x_p)=\frac{1-p}{1+p}-\Omega.\label{htpoint}
\end{equation}
Thus, with both an extended and a point defect, the location of the LDW is at the middle of CHI.
Again as for an extended defect, this is a consequence of the symmetry in the forms of $n_{L}(x)$ and $n_{R}(x)$.
For $x_L=x_R$, the extent of the MC phase vanishes and one obtains a straight line smoothly connecting the densities $n_{L}$ and $n_{R}$.
Our MFT results here are complemented by extensive MCS studies.
Plots of $ n_I(x)$ versus $x$ for various $p$ and $\Omega$ in the steady states are shown in
Figs.~\ref{point_wall}, \ref{st_point} and \ref{point_3phase}. As for an extended defect, we find
the steady state average density $n=1/2$ in the steady state by using the MF form of
$n_I(x)$ as given in Eq.(\ref{pointmf}), again in agreement with the corresponding MCS results. Lastly, there are no pure LD, HD and MC phases in CHI for reasons very similar to the reasons for nonexistence of those phases in CHI with an extended defect, as discussed above.


\begin{figure}[htb]
\includegraphics[height=6.5cm]{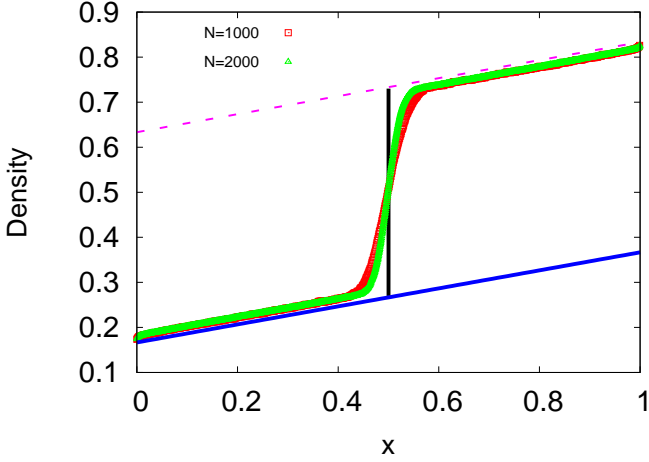}
\caption{(Color online) Density plot for $x_L>x_R$. $\Omega=0.2, p=0.2$. The square points
and the triangular points show MCS results for $N=1000$ and $N=2000$, respectively.
The black line corresponds to the position of DW according to MFT.The gapped and solid lines
represent $n_R$ and $n_L$ MFT equations, respectively. It can be seen that the
DW becomes sharper on increasing the system size.} \label{point_wall}
\end{figure}

\begin{figure}[htb]
\includegraphics[height=6.5cm]{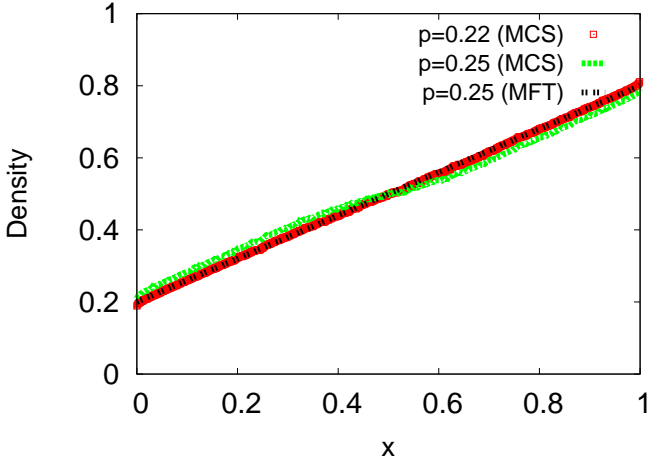}
\caption{(Color online) Straight line profile for the
point defect ($x_L=x_R, N=1000$). Even though qualitatively MFT and MCS agree,
minor quantitative disagreements between the two for the
point defect case can be seen. The dotted black line represents
the MFT density profile equation for $\Omega=0.6,p=0.25$ and the green line
shows the corresponding MCS result. The red points correspond to the
MCS density profile for $\Omega=0.6,p=0.22$.} \label{st_point}
\end{figure}

\begin{figure}[htb]
\includegraphics[height=6.5cm]{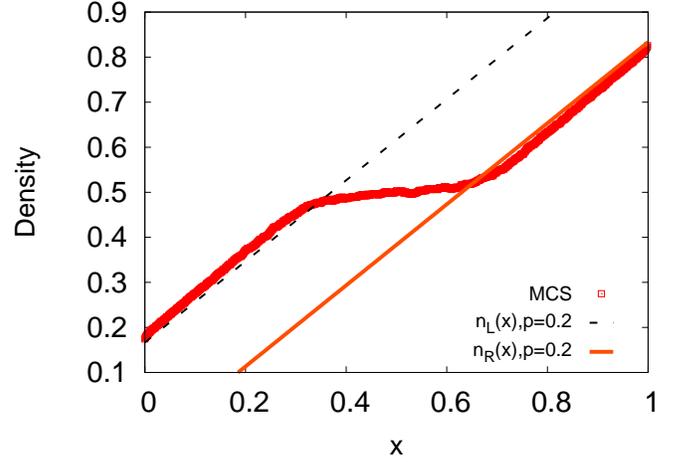}
\caption{(Color online) Density plot for three-phase coexistence for
the point defect. $\Omega=0.9, p=0.2, N=1000$. The gapped and
solid lines represent MFT equations while the square points
are obtained from MCS.} \label{point_3phase}
\end{figure}

\subsection{Phase diagram for a point defect}

To construct the phase diagram in the $\Omega-p$ plane, we identify the threshold line for crossover
from three-phase to two-phase coexistence. The crossover line is obtained by the condition
$x_L=x_R$. This yields
\begin{equation}\label{point_boundary}
 p=\frac{1-\Omega}{1+\Omega}.
\end{equation}
The phase boundary is shown in Fig.~\ref{phase_diagram_point}.
Following our analysis above for an extended defect, we consider the width of the MC phase,
$W_p=x_R-x_L>0$ as the order parameter for the phase transition
between the three- and two-phase coexistences; $ W_p$ is zero in the two-phase coexistence.
Again as above, $W_p$ depends linearly on $\delta p_p$ and
$\delta \Omega_p$, where $p=p_{cp}+\delta p_p,\Omega=\Omega_{cp}-\delta \Omega_{p}$;
$\Omega_{cp}$ and $p_{cp}$ satisfy Eq.~(\ref{point_boundary}). Clearly,
$ W_p$ vanishes smoothly as $\delta p_p$ and $\delta\Omega_p$, indicating
the second order nature of the phase transition. Again, the corresponding order
parameter exponent is unity, same as that for an extended defect.

\begin{figure}[htb]
\includegraphics[width=8.5cm]{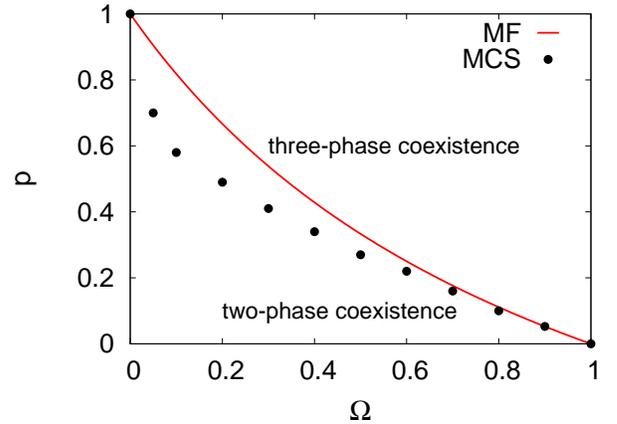}
\caption{(Color online) The phase diagram for a point defect. The curve
$p=(1-\Omega)/(1+\Omega)$ separates regions of two-phase
and three-phase coexistences.} \label{phase_diagram_point}
\end{figure}

\section{Comparison between the density profiles with extended and point defects}\label{comparison}

We find that both with extended and point defects, the system can
be either in three-phase or two-phase coexistence. As a result, the phase diagrams (\ref{phase_diagram})
and (\ref{phase_diagram_point}) respectively, for extended and
point  defects have  similar structures. 
However, the precise phase boundaries in the two cases differ significantly quantitatively.
Similar differences are observed in plots of density profiles with both extended and point
defects; for plots of density versus $x$, see Figs.~\ref{mismatch1} and~\ref{mismatch2} for point defects and Fig.~\ref{match} for
extended defect.
 The difference between MFT and
MCS results is markedly visible in Fig.~\ref{mismatch2}, i.e.,
for sufficiently small $\Omega$. Fig. 9 shows that for certain range
of values of the phase parameters $p$ and $\Omega$, MFT and MCS
yield contrasting results. For example,  density
profiles for a point defect  ($p=0.6, \Omega=0.2$) are shown in Fig.~\ref{mismatch2}.
These  chosen values of $p$ and $\Omega$ are such that they lie in
the region of the
phase space where MFT and MCS phase boundaries do not overlap (see
Fig.~\ref{phase_diagram_point}). These disagreements are clearly visible in Fig.~\ref{mismatch2}. While the
thick continuous lines (with LDW at $x=0.5$) in Fig.~\ref{mismatch2} correspond to MFT solutions for the
two-phase coexistence with $p=0.6, \Omega=0.2$ (see Eqs.~\ref{pointmf}) , the MCS
studies on
the other hand, show three-phase coexistence for the same $p$ and
$\Omega$ (data points in Fig.~\ref{mismatch2}). The MCS results for 
$N=1000, 2000, 3000$ show no systematic system
size dependences
and overlap with each other. Notice that the relative quantitative
inaccuracies of MFT in a
closed ring TASEP with a point defect without any LK, in comparison
with a closed
TASEP with extended defects (again without LK) have been known; see,
e.g. Ref.~\cite{niladri}
and Ref.~\cite{tirtha1} for studies on closed TASEPS with point and
extended defects, respectively.
It is, perhaps, not a surprise that for small $\Omega$, some of theses
mismatches are observed in our model as well.

\begin{figure}[htb]
\includegraphics[width=8.5cm]{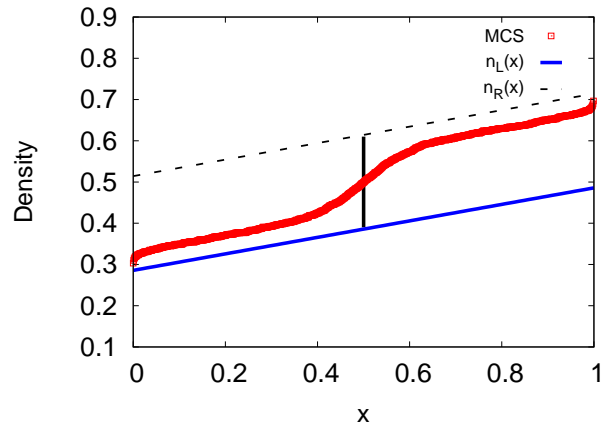}
\caption{(Color online) Plot of the density profile for a point defect displaying the mismatch between MFT and MCS predictions ($\Omega=0.2$, $p=0.4$, $N=1000$).
The black line at $x=0.5$ corresponds to the MFT LDW. Gapped and solid straight lines
refer to MF equations for $n_R(x)$ and $n_L(x)$, respectively. Square points show MCS results.} \label{mismatch1}
\end{figure}

\begin{figure}[htb]
\includegraphics[width=8.5cm]{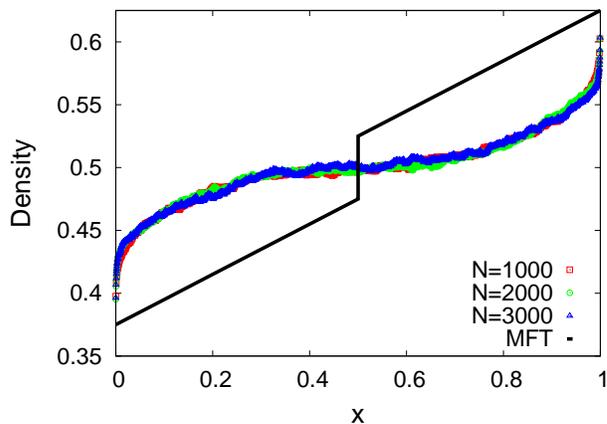}
\caption{(Color online) Mismatch  between MFT and MCS predictions for the density profile with a
 point defect  ($\Omega=0.2$, $p=0.6$, $N=1000$ (square), $2000$ (circle), $3000$ (triangle)).
 While the MFT solutions, shown by the continuous black lines (with LDW at
 $x=0.5$) for the given values of $\Omega$ and $p$ display a two-phase coexistence, MCS data 
points for all the three values of $N$ display a
three-phase coexistence. This disagreement is also visible in 
Fig.~\ref{phase_diagram_point}, 
where the MCS and MFT phase boundaries differ substantially for small $\Omega$. 
} \label{mismatch2}
\end{figure}

\begin{figure}[htb]
\includegraphics[width=8.5cm]{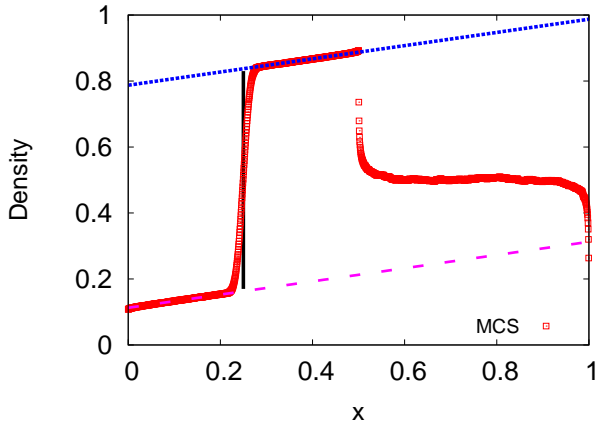}
\caption{(Color online) Density profile for the system with an
extended defect ($\Omega=0.2$, $p=0.4$, $N=2000$) showing strong agreement between MCS and MFT predictions.
The black vertical line corresponds to the MFT LDW. Thin gapped($n_{1\alpha}(x)$)
and thick broken($n_{1\beta}(x)$) lines refer to the MF equations. The square points
show the MCS results.}\label{match}
\end{figure}

These differences may be understood in terms of
the differences in the steady state currents across the defects in the two cases.
In case of an extended defect, the current in the neighborhood of the junctions A and B is $J_e=p/4$,
where as it is $J_p=p/(1+p)^2$ very close to a point defect. Thus, $J_e < J_p$ generically, since $p<1$.
Now, the existence of a three-phase coexistence requires the currents rising from low values ($J_e$ or $J_p$,
as the case may be) to 1/4. The corresponding densities in CHI also vary {\em linearly} with slope $\Omega$ from low or high values near the extended or point defects
to reach 1/2 at the meeting points with the MC part of the three-phase coexistence, i.e., at $x_\alpha$ and $x_\beta$ for an extended defect and $x_L$ and $x_R$ for a point defect. Since $J_e<J_{p}$ generically,
the corresponding values of $n_I(x)$ near a point defect is {\em closer} to 1/2 than they are near the junctions A and B with an extended defect: $n_{I\alpha}(x=0) < n_L(x=0)$ and $n_{I\beta} (x=l)>n_R(x=1)$ for a given $p$.
 Now, for a given $\Omega$, the slope of the spatially varying solutions for $n_I(x)$ are same ($\Omega$) for both extended and point defects. The densities $n_{I\alpha}(x),n_{I\beta}(x)$ (for an extended defect)
 and $n_L(x),n_R(x)$ (for a point defect) must reach 1/2 in the bulk for the MC phase to exist. Hence,
 with an extended defect for a given $p$
 higher values of $\Omega$ are required for the system to reach the threshold of existence
for the MC part of the three-phase coexistence. This explains the quantitative differences in the phase boundaries in the phase diagrams (\ref{phase_diagram}) and (\ref{phase_diagram_point}).

Notice that for the phase boundary
corresponding to an extended defect (Fig.~\ref{phase_diagram}), the level of quantitative agreement between the MFT and
MCS results for small-$\Omega$ is much stronger than in the phase diagram (Fig.~\ref{phase_diagram_point})
for a point defect. Similarly, density profiles for a point defect show much larger discrepancy between the MCS and the MFT results, in comparison with the same for an extended defect;
see Figs.~\ref{mismatch1} and \ref{mismatch2} for a point defect which clearly display the quantitative disagreement
between MCS and MFT results. In fact, this disagreement is revealed vividly in Fig.~\ref{mismatch2},
where the density profile obtained from MCS describes a three-phase coexistence, where as MFT predicts a
two-phase coexistence. In contrast, the density profile for CHI with an extended defect as obtained from MCS
match very well with the corresponding MFT prediction, see Fig.~\ref{match}. These features may be qualitatively
understood as follows. In the small-$\Omega$ limit, the effects of particle
nonconservation is small, particle number is weakly conserved and hence the correlation effects
due to (weak) conservation of particle number should be substantial, rendering MFT quantitatively inaccurate.
For an extended defect with $0<l<1$, the total particle number in CHI should fluctuate within a range $l/2\pm(1-l)/2$, assuming $l>1-l$ and the averaging occupation number to be 1/2 in CHI and CHII (borne out by our MCS and MFT).
 Thus, even in the small-$\Omega$ limit, particle number fluctuations in CHI is expected to be
still substantial with an extended defect, weakening the correlation effects. This explains the
good agreement between the MFT and MCS
for extended defects. In contrast, for a point defect the  total particle number fluctuations in CHI
should vanish in the small-$\Omega$ limit, since $l\rightarrow 1$ or CHII has a vanishingly small size.
Hence, the correlation effects should be
large, causing larger discrepancies between the MCS and MFT results for a point defect in the small-$\Omega$ limit.
For larger $\Omega$, LK ensures lack of any correlation effect regardless of a point or an extended defect,
leading to good agreements between MFT and MCS results for both of them.


\section{Why $n=1/2$ for all $p$ and $\Omega$?}\label{smallomega}

As we have shown above,
for both point and extended defects the system is half-filled in the steady state, i.e., the global average density $n=1/2$  for all $\Omega$ and $p$, disregarding
the possible boundary layers (which have vanishing thickness in the thermodynamic limit). This result may be
easily obtained by using the MF form of $n_I(x)$ (and also of $n_{II}(x)$ in case of an extended defect) as given
above for point and extended defects; our MCS results also validate this to a good accuracy. That $n=1/2$ generically
in our model may be understood physically as follows. Notice that if we set $p=1$ (homogeneous limit), the model is
translationally invariant and the average steady state density {\em at every site} is 1/2. Thus, there are equal
number of particles and holes on average. A shift of $n$ from 1/2 would indicate either more particles or
more holes in the system in the steady state. However, even when inhomogeneity is introduced ($p<1$),
there is nothing that favors either particles or holes, since the inhomogeneity that acts as an inhibitor
for the particle current, also acts as an inhibitor for the holes equally. Thus, it is expected to have equal
number of particles and holes on average, i.e., $n=1/2$ even with $p<1$. Notice that this argument does not
preclude any local excess of particles or holes, since the particles and the holes move in the opposite
directions, and hence the presence of a defect should lead to excess particles on one side and excess holes
(equivalently deficit particles) on the other side of it. This holds for any $\Omega$,
including $\Omega\rightarrow 0$,  and rules out a pure LD or a pure HD phase. This is in contrast
with coupled TASEP and LK with open boundaries~\cite{Frey-LK}, where the boundary conditions
explicitly favor more particles or holes, and consequently make pure
LD or HD phases possible; in the limit of small-$\Omega$, the boundary conditions dominate and the
density profiles smoothly crossover to those of pure open TASEP. In contrast, in the present model,
the density profiles do not smoothly crossover to those for a closed TASEP (with uniform density) for any small-$\Omega$. Instead however, for $\Omega\rightarrow 0$, the LDW heights (\ref{htextnd}) and (\ref{htpoint}) respectively for extended and point defects with LK smoothly reduce to the results of Ref.~\cite{Mustansir1} and Ref.~\cite{lebo}, respectively with $n=1/2$.

\section{Summary and outlook}\label{conclu}

In this article, we have constructed an asymmetric exclusion process on an inhomogeneous ring
with LK. Our MFT and MCS results reveal that in the limit when TASEP along the ring competes with LK,
the model displays inhomogeneous steady state density profiles and phase transitions between different
states of coexistence, parametrized by the scaled LK rate $\Omega$ and hopping rate $p<1$ at the defect site(s).
The model always has a mean density $n=1/2$ for all $\Omega$ and $p$; as a result there are no pure LD or
HD phases, in contrast to LK-TASEP with open boundaries. Our model may be extended in various ways. For
instance, we may consider unequal attachment ($\omega_A$) and detachment $\omega_D$ rates ($\omega_A/\omega_D=K\neq 1$). In this case, by using the
arguments above, $n=K/(1+K)$, see ~\cite{Frey-LK} and can be more or less than 1/2. Thus, far more complex phase behavior including
spatially varying LD or HD phases should follow~\cite{Frey-LK}. Details will be discussed elsewhere. Additionally, one may introduce multiple defect
segments or point defects. However, unlike Refs.~\cite{niladri,tirtha1}, no delocalized domain walls (DDW) are
expected even when various defect segments or point defects have same hopping rates. This is because DDWs in
Refs.~\cite{niladri,tirtha1} are essentially consequences of the strict particle number conservations in the models
there, the latter being absent in the presence of LK. Our results very amply emphasize the relevance of the ring or closed geometry of the system in the presence of LK. The simplicity of our model limits direct applications
of our results to practical or experimental situations. Nonetheless, our results in the context of traffic along
a circular track with constrictions or ribosome translocations along mRNA loops with defects, together with
random attachments or detachments, generally show that the steady state densities should be {\em generically inhomogeneous} regardless of the
details of the defects. We hope experiments on ribosomes using ribosome profiling techniques~\cite{ribopro} and numerical
simulations of more detailed traffic models should qualitatively validate our results.

A few technical comments are in order. First of all, notice that our MFT analysis is equivalent to considering CHI
as an open TASEP with selfconsistently obtained injection ($\alpha_e/\alpha_p$) and extraction ($\beta_e/\beta_p$) rates,
with $\alpha_e=\beta_e\leq 1/2$ and $\alpha_p=\beta_p\leq 1/2$, as obtained above. Given this analogy with an open TASEP,
we can now compare our results with that of Ref.~\cite{Frey-LK}, which has investigated an open TASEP with LK.
With the conditions on $\alpha_{e,p}$ and $\beta_{e,p}$, our results should correspond to the the $\alpha=\beta\leq 1/2$ line
in the phase diagrams of Ref.~\cite{Frey-LK}, where $\alpha$ and $\beta$ are the injection and extraction rates in Ref.~\cite{Frey-LK}.
Now notice that in Ref.~\cite{Frey-LK} along the line $\alpha=\beta\leq 1/2$ only two-phase and three-phase coexistences are possible,
in agreement with our results here. Furthermore, as in Ref.~\cite{Frey-LK}, two-phase coexistence is found for small $\alpha,\beta$,
which in our model means small $p$, where as for large $p$, three-phase coexistence is found. Our MFT is based on analyzing the bulk density profile,
neglecting the boundary layers. An alternative powerful theoretical approach to TASEP-like models with open boundaries have been formulated
that makes use of the boundary layer itself, instead of the bulk~\cite{somen}. It will be interesting to extend these ideas to TASEP on a closed ring with LK.

\section{Acknowledgement}
One of the authors (AB) gratefully acknowledges partial financial support by
the Max-Planck-Gesellschaft (Germany) and Indo-German Science \& Technology
Centre (India) through the Partner Group programme (2009). A.K.C. acknowledges the financial support from
C.S.I.R. under the Senior Research Associateship [No. 13(8745-A)/2015-Pool].

\end{document}